\documentclass[preprint]{aastex63}

\shorttitle{A signature of tides in the WASP-12 system}
\shortauthors{Maciejewski et al.}

\begin{document}

\title{An apparently eccentric orbit of the exoplanet WASP-12~b as a radial velocity signature of planetary-induced tides in the host star}

\correspondingauthor{Gracjan Maciejewski}
\email{gmac@umk.pl}

\author[0000-0002-4195-5781]{Gracjan Maciejewski}
\affiliation{Institute of Astronomy, Faculty of Physics, Astronomy and Informatics, Nicolaus Copernicus University, ul.~Grudziadzka 5, 87-100 Toru\'n, Poland}

\author[0000-0002-0587-8854]{Andrzej Niedzielski}
\affiliation{Institute of Astronomy, Faculty of Physics, Astronomy and Informatics, Nicolaus Copernicus University, ul.~Grudziadzka 5, 87-100 Toru\'n, Poland}

\author[0000-0003-4936-9418]{Eva Villaver}
\affiliation{Departamento de F\'isica Te\'orica, Universidad Aut\'onoma de Madrid, Cantoblanco 28049 Madrid, Spain}

\author{Maciej Konacki}
\affiliation{Nicolaus Copernicus Astronomical Center, Department of Astrophysics, ul.~Rabia\'nska 8, 87-100 Toru\'n, Poland}

\author{Rafa\l{} K. Paw\l{}aszek}
\affiliation{Nicolaus Copernicus Astronomical Center, Department of Astrophysics, ul.~Rabia\'nska 8, 87-100 Toru\'n, Poland}

\begin{abstract}
Massive exoplanets on extremely tight orbits, such as WASP-12 b, induce equilibrium tides in their host stars. Following the orbital motion of the planet, the tidal fluid flow in the star can be detected with the radial velocity method. Its signature manifests as the second harmonics of the orbital frequency that mimics a non-zero orbital eccentricity. Using the new radial velocity measurements acquired with the HARPS-N spectrograph at the Telescopio Nazionale Galileo and combining them with the literature data, we show that the apparent eccentricity of WASP-12 b's orbit is non-zero at a 5.8 sigma level, and the longitude of periastron of this apparently eccentric orbit is close to 270 degrees. This orbital configuration is compatible with a model composed of a circular orbit and a signature of tides raised in the host star. The radial velocity amplitude of those tides was found to be consistent with the equilibrium tide approximation. The tidal deformation is predicted to produce a flux modulation with an amplitude of 80 ppm which could be detected using space-born facilities.
\end{abstract}



\section{Introduction}\label{Sect.Intro}

The WASP-12 system belongs to a small group of planetary systems with giant planets on extremely tight orbits. The late F-type host star is orbited by the bloated hot Jupiter WASP-12~b with an orbital period $P_{\rm{orb}}$ of about 1.09 d \citep{2009ApJ...693.1920H}. The proximity of the host star (i.e., 0.023 a.u. or about 3 stellar radii) results in an equilibrium temperature of the order of 2500 K. This unique system architecture has given rise to a number of studies on the planetary atmosphere and planet-star interactions \citep[see][for a comprehensive review]{2017haex.bookE..97H}. The planet was found to be surrounded by a translucent exosphere producing strong absorption by resonance lines of metals in the near-UV \citep{2010ApJ...714L.222F}. The exospheric gas overfills the Roche lobe and the planet is loosing mass via both the Lagrangian L1 and L2 points. Numerical simulations show that the gaseous envelope forms a circumstellar disk \citep{2018MNRAS.478.2592D}. 

\citet{2016A&A...588L...6M} detected apparent shortening of the orbital period that could be caused by shrinking of the orbit due to tidal decay or could be a part of the long-term periodic variations produced by apsidal precession. Apsidal precession was found to be disfavoured by new transit and occultation timing \citep{2017AJ....154....4P,2018AcA....68..371M,2019arXiv191109131Y} and gives an upper limit for the orbital eccentricity $e_{\rm{b}}$ of the order of $10^{-3}$ \citep{2016A&A...588L...6M,2017AJ....154....4P}. Such a small value is not surprising because the planetary orbit is expected to be circularised on a relative short timescale due to efficient dissipation of planetary tides. The rate of the tidal decay is related to the modified tidal quality factor $Q'_{*}$, which parametrises the response of the star's interior to tidal perturbation induced by a planet. For the WASP-12 system, the value of $Q'_{*}$ was found to be of the order of $10^5$ \citep{2016A&A...588L...6M,2017AJ....154....4P,2018AcA....68..371M,2019arXiv191109131Y} that is 1--2 orders of magnitude lower than the typical values obtained from studies of binary stars \citep[e.g.][]{2005ApJ...620..970M} and other planetary systems \citep[e.g.][]{2017A&A...602A.107B}. As discussed by \citet{2019MNRAS.482.1872B}, the nature of this discrepancy remains unresolved.

Using the equilibrium tide approximation, \citet{2012MNRAS.422.1761A} showed that tides, which are risen by a massive planet in its host star, could be detected with the radial velocity (RV) method. These tidal deformations of the star are expected to manifest themselves in the form of a RV signal with an amplitude of a few m~s$^{-1}$. The period of this signal is half of the orbital period and its phase is related to the planetary orbital motion in such a way that the RV signature of tides can be mimicked by an apparently non-zero orbital eccentricity and a longitude of periastron equal to 270$\degr$. In this study, we demonstrate that these conditions are met in the WASP-12 system.

\section{Observational data}\label{Sect.Obs}

\subsection{New RV observations}\label{SubS.NewObs}

We acquired 17 RV measurements with the High Accuracy Radial velocity Planet Searcher in the northern hemisphere \citep[HARPS-N,][]{2012SPIE.8446E..1VC} fed by the 3.58 m Telescopio Nazionale Galileo (TNG), located at the Observatorio del Roque de los Muchachos on La Palma (Spain). The instrument is an echelle spectrograph covering the wavelength range between 383 nm and 693 nm with a maximal resolving power of $R=115000$. Spectra were gathered between 2013 January 02 and 2017 November 16, most of them as a backup of the Tracking Advance Planetary Systems (TAPAS) project \citep{2015A&A...573A..36N,2017A&A...606A..38V}. The standard user pipeline, which is based on the weighted cross-correlation function method, was used to reduce the data and to determine the high precision RV measurements and their uncertainties. The simultaneous Th-Ar calibration mode of the spectrograph was used for wavelength calibration. The G2 cross-correlation mask, which is the closest to the spectral type of WASP-12, was used to determine RVs. The details on individual observations are given in Table~\ref{tab:RvObs}.

\begin{table*}
 \centering
 \caption{Individual Doppler observations. UT start is the date of the beginning of the exposure. $t_{\rm{exp}}$ is the exposure time. $X$ shows the airmass change during the exposure. $d_{\rm{Moon}}$ is the angular distance of the Moon at the middle of the exposure. $\rm{BJD}_{\rm{TDB}}$ is barycentric Julian date in barycentric dynamical time of the exposure centroid. RV and $\sigma_{\rm{RV}}$ are the determined value of radial velocity and its uncertainty, respectively.}
 \label{tab:RvObs}
 \begin{tabular}{l c c c c c c} 
 \hline
UT start  & $t_{\rm{exp}}$ (s)  & $X$ & $d_{\rm{Moon}}$ ($\degr$) & $\rm{BJD}_{\rm{TDB}}$ & RV (km s$^{-1}$) & $\sigma_{\rm{RV}}$ (km s$^{-1}$)\\
 \hline
2013 Jan 02, 02:54:44 & 1605 & $1.14 \rightarrow 1.22$ & $ 59.8$ & 2456294.637014 & 18.9829 & 0.0028 \\
2013 Jan 28, 23:59:50 & 1458 & $1.03 \rightarrow 1.06$ & $ 55.3$ & 2456321.513859 & 19.3190 & 0.0033 \\
2013 Mar 22, 22:58:20 & 1949 & $1.44 \rightarrow 1.66$ & $ 33.6$ & 2456374.469642 & 18.8632 & 0.0033 \\
2013 Apr 28, 21:10:47 & 2202 & $1.73 \rightarrow 2.14$ & $162.6$ & 2456411.392928 & 19.0165 & 0.0031 \\
2013 Dec 09, 03:03:08 & 1500 & $1.01 \rightarrow 1.04$ & $115.9$ & 2456635.641731 & 19.1406 & 0.0105 \\
2013 Dec 20, 23:29:26 & 1288 & $1.12 \rightarrow 1.08$ & $ 34.3$ & 2456647.492182 & 18.9412 & 0.0055 \\
2013 Dec 21, 04:46:35 & 1442 & $1.35 \rightarrow 1.47$ & $ 35.9$ & 2456647.713618 & 19.2107 & 0.0044 \\
2014 Jan 27, 19:33:06 & 1605 & $1.42 \rightarrow 1.29$ & $163.6$ & 2456685.329266 & 19.0232 & 0.0095 \\
2014 Jan 28, 00:35:28 & 1481 & $1.07 \rightarrow 1.11$ & $166.4$ & 2456685.539009 & 18.8710 & 0.0045 \\
2014 Mar 23, 21:29:52 & 1904 & $1.13 \rightarrow 1.21$ & $168.0$ & 2456740.407717 & 19.1097 & 0.0033 \\
2014 Apr 08, 20:56:10 & 1800 & $1.20 \rightarrow 1.31$ & $ 31.1$ & 2456756.381790 & 18.9077 & 0.0040 \\
2014 Apr 09, 22:02:54 & 1800 & $1.51 \rightarrow 1.74$ & $ 42.7$ & 2456757.427242 & 18.9604 & 0.0161 \\
2014 Apr 11, 21:38:47 & 1800 & $1.43 \rightarrow 1.62$ & $ 66.3$ & 2456759.411696 & 19.1810 & 0.0073 \\
2014 Apr 22, 20:51:41 & 1852 & $1.34 \rightarrow 1.50$ & $144.9$ & 2456770.367480 & 19.1278 & 0.0024 \\
2015 Feb 12, 23:01:28 &  500 & $1.02 \rightarrow 1.02$ & $145.6$ & 2457066.463996 & 18.8624 & 0.0102 \\
2015 Apr 22, 21:00:29 & 2206 & $1.37 \rightarrow 1.59$ & $ 16.5$ & 2457135.373626 & 18.9487 & 0.0037 \\
2017 Nov 16, 03:38:31 & 1187 & $1.01 \rightarrow 1.00$ & $ 110.9$ & 2458073.663121 & 18.9540 & 0.0042 \\
 \hline
 \end{tabular}
\end{table*}

\subsection{Literature data}\label{SubS.Lit}

We used the RV measurements from \citet{2009ApJ...693.1920H} and \citet{2011MNRAS.413.2500H}. They were acquired with the SOPHIE spectrometer \citep{2008SPIE.7014E..0JP} and the 1.9 m telescope at the Observatoire de Haute Provence (France) in the observing seasons 2007/8, 2008/9, and 2009/10. Since \citet{2011MNRAS.413.2500H} note that the velocity zero-point floats by several dozen m~s$^{-1}$ in a timescale of several months, the dataset was split into 3 subsets for the individuals seasons each.

Precise Doppler measurements were extracted from \citet{2014ApJ...785..126K}, including reprocessed observations originally used by \citet{2012ApJ...757...18A}. That survey was performed with the High Resolution Echelle Spectrometer \citep[HIRES,][]{1994SPIE.2198..362V} coupled with the 10 m Keck I telescope between 2009 and 2013.

Additional data obtained with HARPS-N were taken from \citet{2017A&A...602A.107B}. Those precise observations were performed within a framework of the Global Architecture of Planetary Systems (GAPS) Consortium \citep{2016MmSAI..87..141P} between 2012 and 2015.

\subsection{Data preprocessing}\label{SubS.DataProc}

In the dataset acquired with SOPHIE in the observing season 2008/9, three measurements have errors 2-3 times greater than the remaining measurements. They were identified as outliers in our preliminary analysis. As discussed by \citet{2013AA...551A.108M}, those measurements were likely affected by clouds and therefore they were skipped in the final iteration.

Since our procedure does not take the Rossiter-McLaughlin (RM) effect into account and some RV measurements were performed when the planet was transiting, the RM signature was subtracted from those measurements. The appropriate corrections of up to 11.4 m~s$^{-1}$ were calculated using a model of the RM effect obtained by \citet{2012ApJ...757...18A}.

The RV jitter is often equated with RV noise produced by stellar intrinsic variability which is caused by convection motions in the stellar envelope and photospherical inhomogeneities \citep{2005PASP..117..657W} or solar-like acoustic waves \citep{2007CoAst.150..106B}. In practice, it is determined as an additional uncertainty that must be added in quadrature to the RV errors in order to obtain a reduced chi-square statistic of unity for an assumed model. Therefore this quantity may contain not only the physical stellar jitter, but also variations from still-undetected planets and components of instrumental and methodological origin \citep{2010ApJ...725..875I,2019A&A...628A.125M}. The stellar jitter for WASP-12 was found to be equal to $9.1^{+1.8}_{-1.3}$ m~s$^{-1}$ by \citet{2017A&A...602A.107B} who used all RV measurements available then and treated jitter as a free parameter while fitting an orbital solution. We noticed, however, that in the case of WASP-12 the jitter is reduced if lower quality data are iteratively rejected. The value of jitter stabilised at $\sim$7.4 m~s$^{-1}$ for the RV measurements with the errors below 8 m~s$^{-1}$. A single night estimate of the jitter, determined for a high-precision Doppler time series acquired on 2012 Jan 01/02 by \citet{2012ApJ...757...18A}, yields a value of 4.8  m~s$^{-1}$. This is de facto a lower constraint on stellar jitter because it does not account for stellar intrinsic variability in timescales longer than a couple of hours. The jitter value of $\sim$7.4 m~s$^{-1}$ represents variations on timescales of years and is greater than the single night estimate by a factor of $\sim$1.5. This is in line with the finding reported by \citet{2019arXiv191010389B} that the ratio of long and short timescale jitter is $1.5-1.7$ for Gyr-old stars. Considering the above, we used the jitter value of 7.4 m~s$^{-1}$ in further analysis. We note that using the single-night estimate of the jitter or the conservative value from  \citet{2017A&A...602A.107B} does not change our quantitative conclusions.

\section{Results}\label{Sec.Res}

\subsection{Orbital eccentricity}\label{SubS.OrbEcc}

A circular-orbit model is characterised by 8 free parameters: an orbital period $P_{\rm{orb}}$, RV amplitude $K$, mean anomaly for a given epoch $M$, and 5 zero-point RV levels for individual datasets each. In addition, the orbital decay rate was included in the model with the decay rate characterised by the change in the orbital period between succeeding transits $\frac{d P_{\rm{orb}}}{d E} = (-9.67\pm0.73) \times 10^{-10}\quad \rm{days~per~epoch}^{2}$ as refined by \citet{2018AcA....68..371M}. The best-fitting solution was found with the Levenberg-Marquardt algorithm. The uncertainties of the parameters were determined with the bootstrap method using $10^5$ resampled datasets. The minimising procedure results in $(\chi^2_{\rm{RV}})_{\rm{circ}} = 186.0$ at 124 degrees of freedom.

The fitting procedure was repeated for a scenario allowing a non-zero eccentricity. Two additional parameters, the orbital eccentricity $e_{\rm{orb}}$ and longitude of periastron $\omega$, were used to parametrise the shape and orientation of the orbit. The 10 parameter model gives $(\chi^2_{\rm{RV}})_{\rm{ecc}} = 142.5$ at 122 degrees of freedom. 
 
To  compare both models, the Bayesian information criterion (BIC) was calculated for each of them following the form
\begin{equation}
  {\rm{BIC}} = {\chi}^2_{\rm{RV}} + k \ln N,
\end{equation}
where $k$ is the number of fit parameters and $N$ is the number of data points. The criterion favours the eccentric-orbit model ($\rm{BIC}_{\rm{ecc}}=291.9$) over the circular-orbit model ($\rm{BIC}_{\rm{circ}}=304.8$) with a probability ratio of $e^{\Delta \rm{BIC}/2} = 6.2 \times 10^{2}$. We notice that the eccentric-orbit model is favoured over the circular-orbit model even if a higher or lower value of stellar jitter is used. For instance, repeating the procedure with the conservative value of jitter of  9.1 m~s$^{-1}$ \citep{2017A&A...602A.107B}  results in the probability ratio of $ 1.5 \times 10^{2}$.

The best-fitting model gives $e_{\rm{orb}} = 0.035 \pm 0.006$ and $\omega = 270.7\degr \pm 0.6\degr$. This is a 5.8$\sigma$ detection of the non-circular orbit. Its orientation is consistent within a 1.2$\sigma$ level with a specific way that is degenerated with the tidal RV signal.

\subsection{Tidal velocity}\label{SubS.TidalVel}

As shown by \citet{2012MNRAS.422.1761A}, an apparently eccentric orbit may be de facto a sum of the first harmonic of the orbital frequency and the second harmonic associated with the tidal velocity. To construct a  model with the tidal velocity component, the RV datasets were phase folded taking the effect of orbital period shortening into account. The barycentric velocity was subtracted but its contribution to the error budget was taken into account by introducing a parameter $\gamma'$ which allows for corrections of the barycentric velocity. The phased RV signal $V_{\rm{rad}}$ was modelled with the formula 
\begin{equation}
  V_{\rm{rad}} = \gamma' + V_{\rm{orb}} + V_{\rm{tide}},
\end{equation}
where 
\begin{equation}
 V_{\rm{orb}} =  - K_{\rm{orb}} \sin \left( 2 \pi (\phi - \phi_{\rm{0}}) \right)
\end{equation}
is the orbital motion component (the first harmonic of the orbital frequency) and 
\begin{equation}
 V_{\rm{tide}} =  K_{\rm{tide}} \sin \left( 4 \pi (\phi - \phi_{\rm{0}}) \right)
\end{equation}
is the tidal velocity component (the second harmonic of the orbital frequency). The parameters $K_{\rm{orb}}$ and $K_{\rm{tide}}$ are the amplitudes, and $\phi_{\rm{0}}$ is the phase offset. The MCMC algorithm was used to find the best-fitting parameters and their uncertainties. The posterior probability distributions were generated using 100 chains, each of which was $10^4$ trials long after discarding the first 1000 steps. The best-fitting parameters were determined as the median values of marginalised posteriori probability distributions, and 15.9 and 84.1 percentile values of the cumulative distributions were used as 1-$\sigma$ uncertainties.

\begin{figure}[ht!]
\includegraphics[width=0.6\columnwidth]{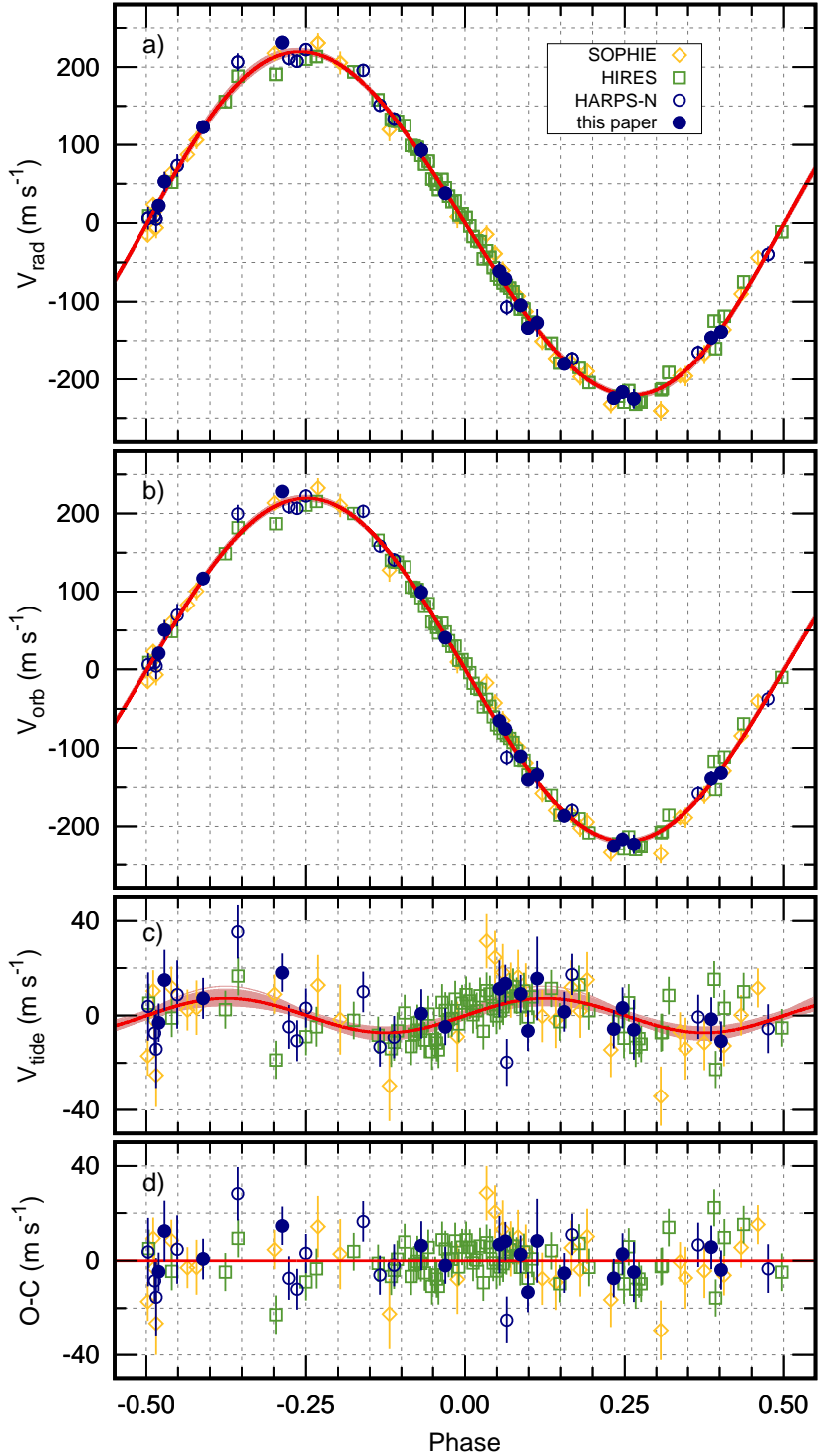}
\caption{\textit{Panel a}: phase-folded RV curve for WASP-12 b with the apparently non-zero eccentricity. Our new observations are marked with dots and the literature data are marked with open symbols. The original errors are increased by the value of jitter of 7.4 m~s$^{-1}$ added in quadrature. The best-fitting model is marked with a red line. The parameter uncertainties of the model are illustrated with pale-red lines which are drawn for 50 sets of parameters, randomly chosen from the Markov chains. \textit{Panel b}: orbital RV component. \textit{Panel c}: tidal RV component which mimics the non-zero orbital eccentricity of WASP-12~b. \textit{Panel d}: the residuals from the best-fitting model. \label{fig:RV}}
\end{figure}

We obtained $K_{\rm{orb}} = 220.0\pm1.3$ m~s$^{-1}$ and $K_{\rm{tide}} = 7.5\pm1.2$ m~s$^{-1}$. The parameters $\gamma'$ with a value of $-0.34 \pm 0.82$ m~s$^{-1}$ and $\phi_{\rm{0}}$ with the value of $(1.0^{+0.7}_{-0.8}) \times 10^{-3}$ were found to be consistent with zero well within 1$\sigma$ and 2$\sigma$, respectively. The best-fitting model together with the orbital and tidal RV components and the residuals, is shown in Fig.\ref{fig:RV}.

\section{Discussion}\label{Sec.Disc}

Preliminary detections of the non-zero eccentricity of WASP-12~b were reported in previous studies with lower significance. In the discovery paper, \citet{2009ApJ...693.1920H} find $e_{\rm{orb}}=0.049\pm0.015$, \citet{2014ApJ...785..126K} report $e_{\rm{orb}}=0.037^{+0.014}_{-0.015}$, and more recently \citet{2019arXiv191109131Y} obtained $e_{\rm{orb}}=0.0317\pm0.0087$. In all those studies, the reported values of $\omega$ were close to $270\degr$, and \citet{2019arXiv191109131Y} pointed out the tidal distortion of the host star as a possible explanation of this specific configuration. On the other hand, \citet{2011MNRAS.413.2500H} find $e_{\rm{orb}}=0.018^{+0.024}_{-0.014}$ that was interpreted as a result speaking in favour of a circular orbit. The same conclusion was reached by \citet{2017A&A...602A.107B} who placed a 1$\sigma$ upper constraint on $e_{\rm{orb}}$ of 0.02.

Following equation~3 in \citet{2006ApJ...649.1004A} and using a conservative value of the planetary quality factor of $10^6$, the circularisation timescale for WASP-12~b is about $0.4$ Myr. This is much shorter than the system age which is estimated to be 4 orders of magnitude longer. If WASP-12's orbital eccentricity of $\sim$0.035 were real, an efficient mechanism that excites and sustains it would be needed to operate in the system. \citet{2019MNRAS.482.1872B} consider perturbations from undetected planetary companion, Kozai-Lidov oscillations, or fluctuations of the gravitational potential induced by stellar convection. First two scenarios may be discarded because no perturbing body has been detected in the system. The magnitude of the third mechanism was found to be negligible. Furthermore, the orbit of WASP-12~b is expected to precess with a period of a few decades. This precession with $e_{\rm{orb}}$ of $\sim$0.035 would produce anti-correlated variations in transit and occultation times with amplitudes of $\sim$20 minutes. No evidence for such scenario was found in timing observations \citep{2017AJ....154....4P,2019arXiv191109131Y}. From this perspective, the tidal fluid flow is a natural explanation for the apparent non-zero eccentricity of WASP-12~b.

In the equilibrium tide approximation, stellar matter is assumed to be incompressible and to follow gravitational equipotentials ignoring fluid inertia. Furthermore, the forcing frequency is set to zero, any effects induced by convective motions are neglected, and the stellar rotation is set to zero. These simplifications make predictions of the equilibrium tide approximation to be accurate to a factor of $\sim$2 \citep{2012MNRAS.422.1761A}. In this context, our determination of the amplitude of the tidal velocity component $K_{\rm{tide}} = 7.3\pm0.8$ m~s$^{-1}$ can be considered as being consistent with the value of 4.78 m~s$^{-1}$ calculated under the equilibrium tide approximation \citep{2012MNRAS.422.1761A}.

The WASP-18 system was identified as the best candidate for detection of the tidal velocity \citep{2012MNRAS.422.1761A}. The amplitude of the tidal RV signal, predicted by the the equilibrium tide approximation, is $\sim$$32$ m~s$^{-1}$. According to \citet{2017A&A...602A.107B}, the orbital eccentricity of WASP-18~b is definitely non-zero with a value of $0.0076\pm0.0010$, and the pericentre longitude of $268.7^{+2.7}_{-2.9}$ degrees agrees with $270\degr$ well within a 1$\sigma$ range. Such configuration corresponds to the tidal RV signal with an amplitude of $\sim$14 m~s$^{-1}$ which, though noticeably smaller, is still not far from the model predictions.

To verify a reliability of our procedure, we reanalysed the data available for the WASP-18 system and compared the outcome to the results reported by \citet{2017A&A...602A.107B}. We used the RV measurements from \citet{2010A&A...524A..25T}, including observations reported by \citet{2009Natur.460.1098H}, and from \citet{2014ApJ...785..126K}. The observations acquired during a transit phase were skipped leaving 53 data points for further analysis -- the same dataset which was analysed in the original study. To place additional constraints on a transit ephemeris, we used all ground-based transit mid-transit times which were published prior to the study of \citet{2017A&A...602A.107B}, as compiled by \citet{2017ApJ...836L..24W}. Following the procedure which we applied to the WASP-12 system, we found that the WASP-18~b's eccentricity is $0.0082\pm0.0010$ and the longitude of periastron is $266.1\degr\pm3.3\degr$. Both quantities agree with the values reported by \citet{2017A&A...602A.107B} well within a 1$\sigma$ range. The parameter uncertainties were found to be comparable with each other that ensures that our procedure does not underestimates uncertainties. This finding strengthens the high detection significance of the non-zero apparent eccentricity for WASP-12~b.

For WASP-12, the tidal amplitude of $7.3$ m~s$^{-1}$ corresponds to the hight of tides up to $\sim$150 km. Such ellipsoidal deformation is predicted to produce a photometric modulation with an amplitude of $\sim$80 ppm. Such signals have been detected in the HAT-P-7 and WASP-18 systems using photometric time series from space-borne telescopes \citep{2010ApJ...713L.145W,2019AJ....157..178S}. Because of relative faintness of the host star ($V=11.7$ mag), the ellipsoidal flux modulation in the WASP-12 system would be possible with such instruments as TESS \citep{2014SPIE.9143E..20R} or CHEOPS \citep{2013EPJWC..4703005B}.

\section{Conclusions}\label{Sec.Conc}

Massive planets on extremely tight orbits induce tidal deformations of their host stars that can be accessible not only by ultra-precise photometric observations, but also by the RV method. We have found that the orbit of WASP-12~b, like the orbit of WASP-18~b, appears to be apparently eccentric with the periastron longitude close to $270\degr$. This is the RV manifestation of the tidal deformation of the host star that follows the orbital motion of the planet. Although the observations are considered as being consistent with predictions of the equilibrium tide approach, development of more advanced models would benefit in our better understanding of planet-star tidal interactions. 

\acknowledgments

We thank the anonymous referee for a prompt and insightful report. GM and AN acknowledge the financial support from the National Science Centre, Poland through grant no. 2016/23/B/ST9/00579. AN is also supported by the National Science Centre, Poland through grant no. 2015/19/B/ST9/02937. EV acknowledges support from the Spanish Ministerio de Ciencia, Innovaci\'on y Universidades under the project  PGC2018-101950-B-I00. MK is supported by the Polish National Science Center (NCN) through grant 2017/27/B/ST9/02727.

\vspace{5mm}
\facilities{TNG(HARPS-N)}

\bibliography{rvTides}{}
\bibliographystyle{aasjournal}

\listofchanges

\end{document}